\documentclass[aps,pre,reprint]{revtex4-1}
\usepackage{epsfig,graphics,amssymb,amsmath,subeqnarray,color}

\def \bnabla{\boldsymbol{\nabla}}

\def \bH{\mathbf{H}}

\def \bR{\mathbf{R}}

\def \bX{\mathbf{X}}

\def \rd{\dot{r}}
\def \Rd{\dot{R}}

\def \zd{\dot{z}}
\def \Zd{\dot{Z}}
\def \fd{\dot{f}}
\def \gd{\dot{g}}

\def \Gd{\dot{G}}
\def \Fd{\dot{F}}

\def \Ef{\mathcal{E}}

\def \Of{\mathcal{O}}

\def \l{\left}
\def \r{\right}

\begin{document}

\title{Buckling instability of squeezed droplets}
\author{Gwynn J. Elfring}
\author{Eric Lauga\footnote{Corresponding author. Email: elauga@ucsd.edu}}
\affiliation{
Department of Mechanical and Aerospace Engineering, 
University of California San Diego,
9500 Gilman Drive, La Jolla CA 92093-0411, USA.}
\date{\today}
\begin{abstract}

Motivated by recent experiments, we consider theoretically the compression of droplets pinned at the bottom on a surface of finite area. We show that if the droplet is sufficiently compressed at the  top by a surface, it will always develop a shape instability at a critical compression. When the top surface is flat, the shape instability occurs precisely when the  apparent contact angle of the droplet at the pinned surface is $\pi$, regardless of the contact angle of the upper surface, reminiscent of past work on liquid bridges and sessile droplets as first observed by Plateau. After the critical compression, the droplet transitions from a symmetric  to an asymmetric shape. The force required to deform the droplet peaks at the critical point then progressively decreases indicative of catastrophic buckling. We characterize the transition
in droplet shape using illustrative examples in two dimensions followed by perturbative analysis as well as numerical simulation in three dimensions. When the upper surface is not flat, the simple apparent contact angle criterion   no longer holds, and a detailed stability analysis is carried out to predict the critical compression.

\end{abstract}
\maketitle

%%%%%%%%%%%%%%%%%%%%%%%%%%%
%%%%%%%%%%%%%%%%%%%%%%%%%%%
\section{Introduction}

The interaction between liquids and solids is ubiquitous in our daily life, from droplets on a windshield to ink in our printers  \cite{degennes04}. Capillary phenomena arise as a consequence of intermolecular forces and  manifest themselves on large scales by the tendency of liquids to  minimize their surface area  \cite{pomeau06}. The theory for the shapes of droplets was proposed over two hundred years ago by Young  \cite{young05} and Laplace  \cite{laplace06}, and since that time much has been learned on the wetting of solids by liquids  \cite{degennes85,quere05}. Of interest to us in this paper is why certain droplet configurations may be unstable. Plateau observed that a liquid jet would ultimately break up into droplets because the energy of the initial cylindrical shape is lowered by long wavelength perturbations  \cite{plateau73}. The dynamics of this instability was later elucidated by Lord Rayleigh  \cite{rayleigh78} while more recent work has explored the finite-time singularity at the break-up  \cite{eggers97}.

A somewhat lesser known observation made by Plateau, in the same work  \cite{plateau73}, details how a droplet suspended between two equal circular disks (a so-called liquid bridge) loses axisymmetry when  sufficiently compressed. Plateau  \cite{plateau73}  observed that the onset of this shape instability occurs when the profile of the droplet at the point of contact with the pinned surface becomes tangent to the disks, hence making for an apparent droplet contact angle of $\pi$  (see also the review in Ref.~\cite{michael81}).  An analytical solution for the shape of a liquid bridge was put forward by Howe \cite{howe87} along with an initial analysis of the instability \cite{howe87,gillette71}. It was later shown formally that if the axisymmetric droplet shape is described in polar coordinates  by a single valued radius function then it is always stable to asymmetric perturbations for pinned boundary conditions  \cite{gillette72}.  Such single-valued description breaks down when the droplet becomes tangent to the disks.  Russo and Steen showed then that past this point the droplet is unstable to asymmetric perturbations \cite{russo86}. 

Subsequent works \cite{meseguer95,lowry95,slobozhanin97,lowry00} further elucidated the space of stability of these liquid bridges for which the results of Plateau is only one of many. In particular, it was similarly shown that a pinned droplet deformed by a gravitational field, rather than by compression from an upper surface, will also transition to an asymmetric shape past the point when the droplet profile is tangent to the pinning line at the point of contact \cite{michael81}. A related shape instability arises when two free-surfaces are squeezed together \cite{bradley01}. If two droplets, or bubbles \cite{bohn03,fortes04}, are brought into contact, then initially the interface separating them  is perpendicular to the direction of compression and their shapes deform in an axisymmetric fashion. Past a critical conformation the separating interface  rotates and the droplets, or bubbles, lose axisymmetry. A general framework for the stability of equilibrium states of capillary phenomena is provided in the text by Myshkis et al. \cite{myshkis87}.

In recent experiments by Nagy and Neitzel, droplets pinned to a bottom surface and compressed by a perfectly non-wetting surface from above have been shown to develop a geometric asymmetry at a critical  deformation  \cite{nagy09}. Experimentally, the perfectly non-wetting condition on the top surface  is obtained using  a thin layer of air  maintained by thermocapillary convection between the cold surface (of arbitrary contact angle) and the hot droplet, leading to an effective  contact angle of $180^\circ$  \cite{neitzel02, aversana04}. On the bottom surface, pinned boundary conditions are achieved by extruding liquid through a capillary, but one can also imagine a droplet confined to the top of a small post or a disk.  When the droplet is compressed, by displacing the top surface, below a threshold the  droplet maintains axisymmetry, but  upon reaching a critical conformation, the droplet  bulges to one side  indicating a shape instability. Subsequent Surface Evolver \cite{brakke92} simulations of the same setup  verified the geometric asymmetry  \cite{nagy09}.

\begin{figure*}
\includegraphics[scale=0.5]{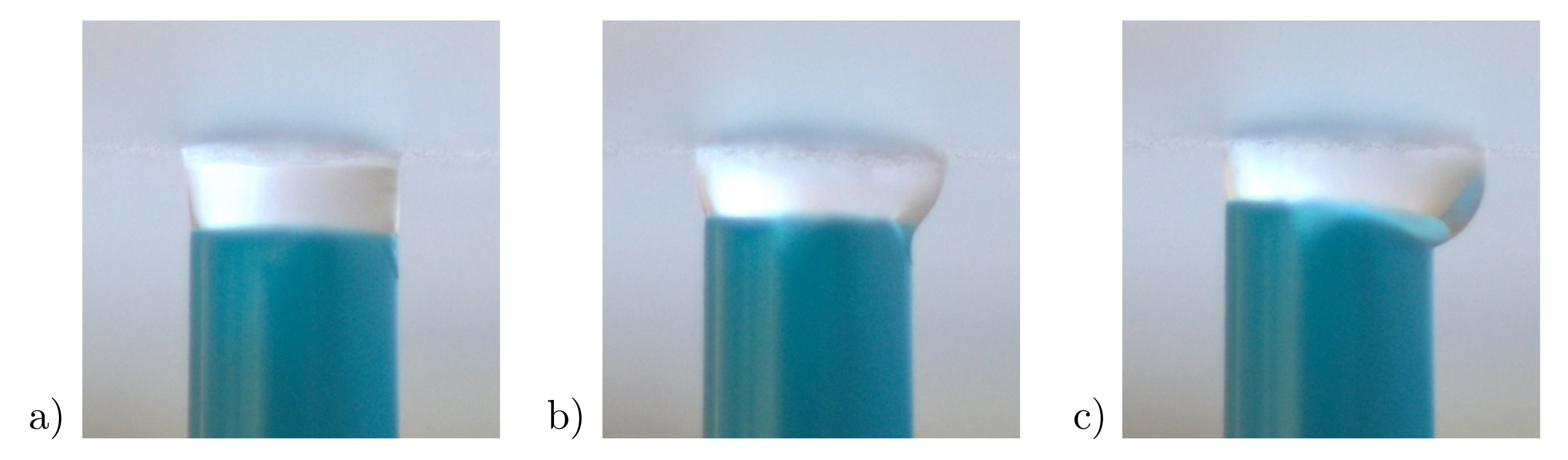}
\caption{Images of a water droplet at the tip of a straw compressed against a hydrophobic surface; (a): The droplet after initial contact, with pinned boundary conditions at the bottom and a fixed hydrophobic contact angle at the top; (b): Axisymmetric deformation of the droplet prior to the onset of shape instability; (c) Asymmetric bulge-like shape of the droplet after the critical compression.}
\label{experiments}
\end{figure*}

In this paper we show that the shape instability observed  by Nagy and Neitzel \cite{nagy09}  arises with the same geometrical criterion as that of the Plateau liquid bridge, namely when the droplet surface becomes tangent to the pinned surface at the point of contact (apparent contact angle of $\pi$). We demonstrate that this instability criterion does not depend on the droplet  contact angle on the upper surface. We reveal, however, that that the geometric instability criterion no longer holds  when the deforming surface is not flat, as is illustrated with a compression of the droplet by a conical surface. We first use a simple table-top experimental example to show that the shape instability occurs even if the deforming surface is not perfectly non-wetting.  We then employ a two-dimensional analysis to provide a preliminary analytical approach to the instability, and derive the stability criterion and its independence on the top contact angle. We next carry out a three-dimensional perturbation energetic analysis confirming the two-dimensional theory, and demonstrating the breakdown of the geometric stability criterion when the deforming surface is no longer flat. We finally utilize numerical computations using Surface Evolver to confirm our theoretical predictions.

%%%%%%%%%%%%%%%%%%%%%%%%%%%
%%%%%%%%%%%%%%%%%%%%%%%%%%%
\section{Table-top experiments}

We first performed simple table-top experiments in an effort to observe whether the shape instability occurs in the case where the compressing surface is not perfectly non-wetting. The results are shown in  Fig.~\ref{experiments}. We use a  plastic straw extruding water, coated with hydrophobic spray to delay its wetting prior to the shape instability.  The water droplet is put in contact with a slightly hydrophobic surface. In Fig.~\ref{experiments} we see that initially the droplet is symmetric (a), and  remains symmetric under small deformations (b). When the droplet has been sufficiently squeezed, it rapidly develops a pronounced geometric asymmetry (c).

%%%%%%%%%%%%%%%%%%%%%%%%%%%
%%%%%%%%%%%%%%%%%%%%%%%%%%%
\section{Energetic analysis}\label{fundamentals}

Consider a droplet  pinned along one surface, which we will call the bottom surface, along a constant circular area. The droplet has an volume $V$ which  remains constant  by conservation of mass. We assume that the droplet is smaller than the capillary length and hence ignore gravity  \cite{degennes04}.  The droplet shape  is the one which minimizes the functional
\begin{eqnarray}
\Ef=\int_S \gamma dS-\int_V p dV,
\label{energy}
\end{eqnarray}
where $\gamma$ is the interfacial tension coefficient which is integrated over all interfaces $S$ and the pressure $p$ is a Lagrange multiplier enforcing mass conservation. We further assume that the droplet is in contact by a second surface, parallel to the bottom surface, and which we will refer to as the top surface. That surface, of area $W$, does not pin the drop but instead the droplet contact line is free to move along the top surface with the set contact angle $\theta$. We ignore effects arising from contact angle hysteresis. In that case the relevant surface energy now becomes
\begin{align}\label{energy1}
E=\gamma_{sv}(W-S_{sl})+\gamma_{sl}S_{sl}+S_{lv}\gamma_{lv}+const.,
\end{align}
where the subscripts $sv$, $sl$ and $lv$ indicate the solid-vapor, solid-liquid and liquid-vapor interfaces respectively.  The surface energy of the pinned interface is unchanging and thus rolled into the constant term. Using  Young's equation for the equilibrium contact angle $\theta$ on the top surface \cite{degennes04}, Eq.~\eqref{energy1}  may then be recast as
\begin{align}\label{energy2}
E=\gamma_{lv}\left(S_{lv}-S_{sl}\cos\theta\right)+\gamma_{sv}W+const.
\end{align}
Given that $W$ is constant, we  see that finding an equilibrium is tantamount to  minimizing the   projected surface $S=S_{lv}-S_{sl}\cos\theta$. Note that when the solid surface is nonwetting, $\theta=\pi$, and one has thus to  minimize $S=S_{sl}+S_{lv}$. From this point on we neglect the constant terms in Eq.~\eqref{energy2}, and drop the $lv$ subscript for the liquid-vapor interfacial coefficient for simplicity, hence we denote $E=\gamma S$.

%%%%%%%%%
%%%%%%%%%
%%%%%%%%%

\subsection{Two-dimensional analysis}

\begin{figure*}
\includegraphics[scale=0.25]{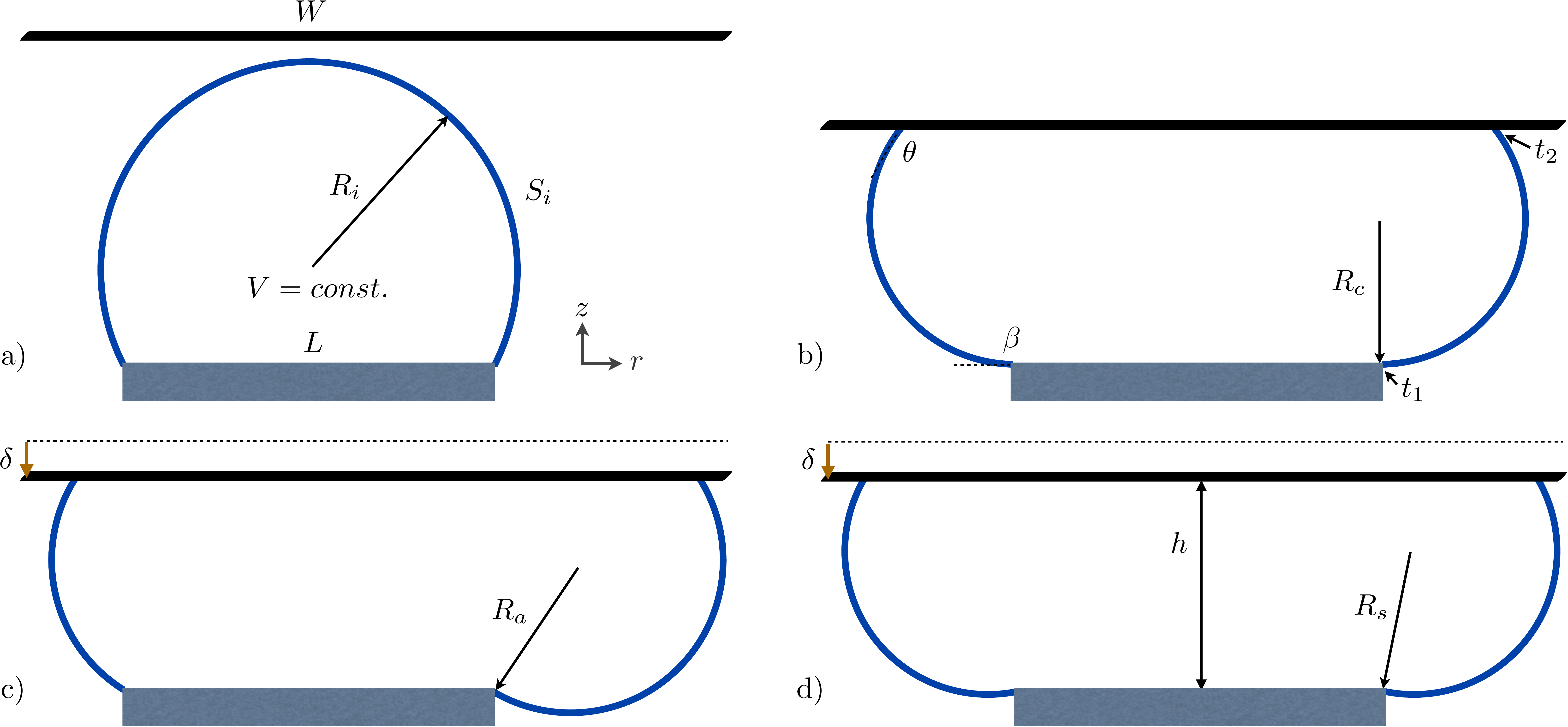}
\caption{Schematic representation of the  droplet in various stages of deformation, for a contact angle  $\theta=2\pi/3$. (a): No deformation; (b): Droplet at the  critical conformation, $R=R_c$;  (c): Asymmetric droplet post threshold $R=R_a$ ($\delta>0$); (d): Symmetric droplet post threshold $R=R_s$ ($\delta>0$). The droplet volume (termed $A$ in 2D, $V$ in 3D)  remains constant; the  initial radius $R_i$ and surface area $S_i$ are thus set by the value of the  droplet volume.}
\label{four}
\end{figure*}

\subsubsection{Flat surface}

In order to get a sense of the essential physics of the droplet instability we first illustrate the process in two dimensions.  The geometric criterion for the instability of symmetric solutions is easier to derive and visualize than in three-dimensions,  which is a calculation we will address  in the next section.  

The setup and notations are illustrated in Fig.~\ref{four}. The volume of the droplet is denoted $V$ in three dimensions and $A$ in two dimensions, and the interfaces are denoted $S$ in both cases.  The liquid-vapor interface which minimizes the total interfacial energy  is a surface of constant mean curvature, and thus in $2D$ it is circular. We therefore write the shape of the liquid-vapor interface, parameterized by $t$ in the \{$r$,$z$\} plane, as
\begin{align}
r(t)&=r_0+R\cos(t),\nonumber\\
z(t)&=z_0+R\sin(t),
\end{align}
where $R$ is the radius of curvature of the 2D droplet originating from the center $(r_0,z_0)$, and  $t$ is the polar angle about that point measured positive counterclockwise from the point where the droplet is locally parallel to the $z$ axis on the right-hand side. The contact point where the droplet is pinned occurs at $t=t_1$, and  the point of contact with the wall $W$ occurs at  $t=t_2$. We define  $\Delta t = t_2-t_1$ so that the surface area of the liquid-vapor portion of the droplet is  $S_{lv}=2R\Delta t$. The contact area with the bottom surface is assumed to be of  length $L=2r_1$. We subsequently employ the subscript $1$ or $2$ on any function of $t$ to denote the function evaluated at $t_1$ or $t_2$ respectively. With this in mind, the height between the two surfaces (clearance) is given by $h=z_2 - z_1$.  We use $\theta$ to denote  the equilibrium contact angle of the droplet on the top surface, and $\beta$ the apparent contact angle with the bottom, horizontal, surface (where the contact line is pinned). 

When the top surface is not in contact with the droplet, as in Fig.~\ref{four}a, then $t_2$ is taken to lie on the axis of symmetry of the undeformed droplet ($t_2=\pi/2$). The initial droplet volume, which  remains constant, is $A=(R_i^2/2)(2\Delta t+\sin[2\pi-2\Delta t])$, and the pinning length $L^2=4R_i^2(1-\sin^2[(\pi-2\Delta t)/2])$ closes the system. The droplet will form a half circle when $L=4\sqrt{A/2\pi}$.

When there is contact with the wall, the droplet takes the shape illustrated in Fig.~\ref{four}b, and two `disconnected' liquid-vapor surfaces are present. In order to maintain mechanical equilibrium these two surfaces must have an identical radius of curvature, $R$. The point $t_2$ is located where the droplet is in contact with the surface, and therefore we necessarily have $t_2=\theta-\pi/2$. If the apparent contact angle at the pinned line, $\beta$, satisfies $\beta \le \pi$, in other words $t_1 \ge -\pi/2$, we will show below that the only solution for the two-dimensional droplet is left-right symmetric (the details below will demonstrate that  there exists no asymmetric solution).

At the critical point (denoted with a subscript $c$), which is depicted in Fig.~\ref{four}b, we have $t_1=t_1^c=-\pi/2$ and $\Delta t=\theta$. The total surface energy is then given by $E_c/\gamma=(2\Theta R_c-L\cos\theta)$, where $\Theta\equiv\theta-\cos\theta\sin\theta$. Here the radius $R_c$ is given by conservation of volume, $A=\Theta R_c^2+LR_c(1-\cos\theta)$; this quadratic equation for $R_c$ can be solved to obtain the critical radius for a given initial droplet volume, pinning length and contact angle. The critical gap height, $h^c=z_2-z_1^c=R_c(1-\cos\theta)$, increases monotonically with the contact angle (which ranges from zero to $\theta=\pi$). 

If we now reduce the gap height by some amount $\delta$, so that $h=h^c-\delta$,  two possibilities exist for the droplet shape, either asymmetric (Fig.~\ref{four}c) or   symmetric (Fig.~\ref{four}d). In each case the left and right sides of the droplet need to have the same radius of curvature to ensure mechanical stability. From this point on all lengths will be non-dimensionalized by the critical radius $R_c$, and hence  $R_c= 1$.

In the asymmetric case,  denoted by the subscript $a$, one of the circular ends moves in and the other moves out (see Fig.~\ref{four}c). The droplet volume in this instance is $A=\Theta R_a^2+hL$, meaning that  $R_a^2=1+L\delta/\Theta$. If we take $\delta<0$ we see that $R_a<1$ which is impossible as the droplet would not span the required length $h$, hence there can be no asymmetric shape for $\delta<0$. Further, we see that for $\delta>0$, $R_a>1$. The relevant surface energy is given by $E_a/\gamma=2\Theta R_a-L\cos\theta$ in that case.

Alternatively, the droplet may remain symmetric  (subscript $s$) past the critical point, as shown in Fig.~\ref{four}d. Conservation of volume leads to  $A=R_s^2 [\Theta -\pi /2- t_1+ \cos t_1(\sin t_1-2\cos \theta)]+hL$, where we set the gap height to be $h=1-\cos\theta-\delta$, and hence $t_1=-\sin^{-1}\l[\cos\theta +h/R_s\r]$. Note that when in contact with hydrophilic surfaces, $\theta<\pi/2$, droplets undergo a inversion of the concavity of the liquid-vapor interface for larger negative values of $\delta$ where this geometry no longer holds. The relevant surface energy is now give by $E_s/\gamma=2(\Theta-t_1-\pi/2)R_s-(L-2R_s\cos t_1)\cos\theta$.

The symmetric and asymmetric cases are the only two possible mechanically stable solutions. The one which appears in equilibrium is the one which minimizes surface energy. Solving for $R_s$ and $E_s$, can in general only be done numerically. However,  the difference in the surface energies between the symmetric and asymmetric configurations, $\Delta E = E_s-E_a$, may be formally bounded as follows
\begin{align}
\frac{\Delta E}{2\gamma R_s}=&\ \Theta+\lambda-\sqrt{\Theta^2+\Theta(\lambda+\Gamma)}\ge 0,
\end{align}
where $\Theta\equiv \theta-\sin\theta\cos\theta$, $\lambda = \alpha-\sin\alpha\cos\theta$ and $\Gamma = (\cos\alpha-\cos\theta)\sin \alpha$ and  while $\theta\in[0,\pi]$ and $\alpha=-(t_1^s+\pi/2)\in[0,\theta]$ (the superscript $s$ indicates the symmetric contact point). We see that evidently $\Theta\ge0$ and $\lambda\ge0$ as $\theta\ge\sin\theta$. Rearranging we obtain\begin{align}
\lambda-\Gamma\ge -\frac{\lambda^2}{\Theta}.
\end{align}
Substituting for back we find
\begin{align}
\alpha-\sin\alpha\cos\alpha\ge-\frac{\lambda^2}{\Theta},
\end{align}
since the left-hand side is non-negative the inequality is proved. The equality holds only at the critical point when $t_1^s=t_1^c=-\pi/2$, in other words when $\delta=0$ and $R_s=1$. We have thus proven that, past the critical point ($\delta>0$), the asymmetric conformation is always the droplet shape which minimizes the free energy, independently of the value of the contact angle on the top surface. Droplets after this critical compression are thus expected to always display an asymmetric shape.

For small values of $\delta$, the value of  $R_s$ can be found by hand, and provided the contact angle $\theta$ is not too small  we obtain
\begin{align}
R_s=&\ 1-\frac{1}{1-\cos\theta} \delta +\frac{(L[1-\cos\theta]+2\Theta)^2 }{256\sin^{10}(\theta/2)} \delta ^2+\Of(\delta^3),\\
\frac{E_s}{\gamma}=&\ 2\Theta -L\cos\theta+L\delta+\frac{L (1-\cos\theta)+ \Theta}{4\sin^4(\theta/2)}\delta^2+\Of(\delta^3).
\end{align}
We then get
\begin{align}
\frac{\Delta E}{\gamma}=\frac{\left(L[1-\cos\theta]+2\Theta\right)^2 }{16\Theta\sin^4(\theta/2)}\delta ^2+\Of(\delta^3)\ge 0.
\end{align}

Notably, the force required to deform the droplet is proportional to the spatial rate of change of the surface area hence, for $\delta > 0$ we find
\begin{align}
f=\frac{\partial E}{\partial \delta}=\gamma\frac{L}{\sqrt{1+L\delta/\Theta}}\cdot
\end{align}
This may be recast in terms of the pressure jump $f/L=\Delta p=\gamma/R_a$ which recovers the Young-Laplace equation in two dimensions.  Since the radius of curvature has increased, the pressure has decreased in this configuration. The slope of the force at the critical point $\l.df/d\delta\r|_{\delta=0}=-\gamma L^2/2\Theta$. As might be physically expected the pressure drop is least steep for large $L$ and $\theta=\pi$.

\begin{figure}
\includegraphics[scale=0.75]{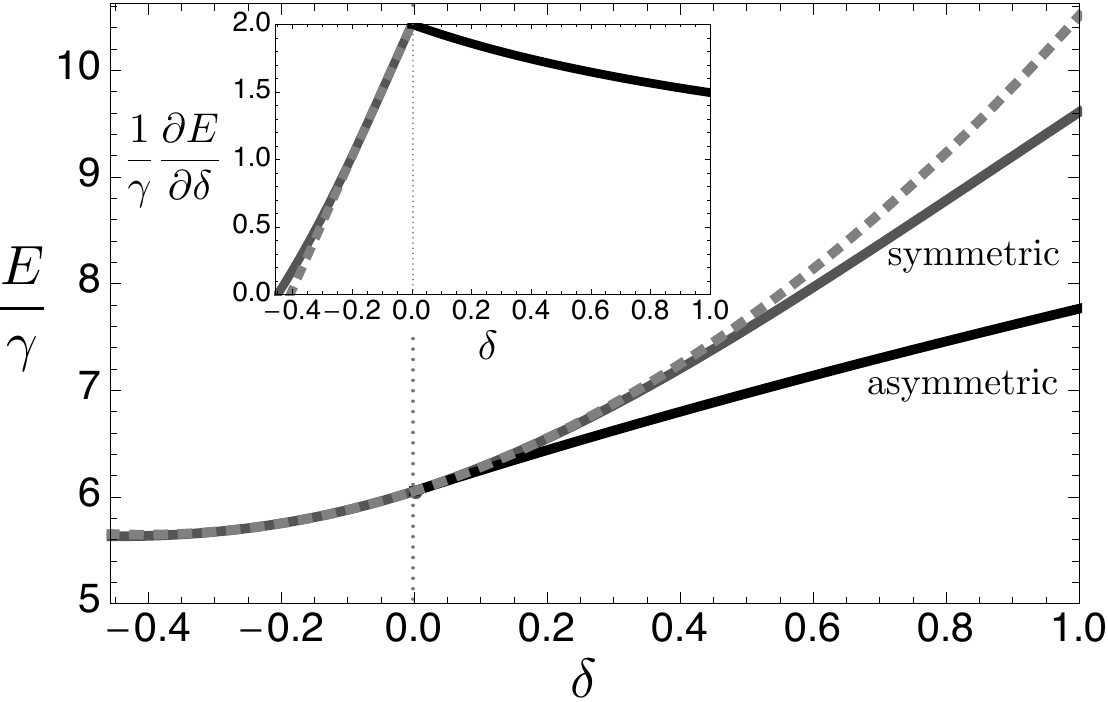}
\caption{Surface energy, $E$, vs.~a change in separation, $\delta$, with $L=2$ and $\theta=2\pi/3$ (two-dimensional case). The solid lines are exact solutions and the dashed line is asymptotic. The inset shows the slope of $E$. We see that when $\delta=0$ (indicated by the dotted line) the surface energy bifurcates into symmetric and asymmetric conformations. The asymmetric conformation (solid black line) has lower surface energy than the asymmetric conformation (solid gray line) and hence is the energetically preferable state.}
\label{energy2d}
\end{figure}

The two-dimensional results are illustrated in  Fig.~\ref{energy2d} where we plot the surface energies, $E$ (main plot), and the energy  slope  (force, inset) against the change in gap height from the critical conformation, $\delta$ (for the values $L=2$ and $\theta=2\pi/3$).  The surface energies increase monotonically as the droplet is progressively compressed.  Beyond the transition point ($\delta=0$) both the asymmetric conformation (black) and the symmetric conformation (gray) are geometrically permissible although the asymmetric conformation is always energetically favorable. Note that the solid gray line is obtained numerically for the symmetric conformation  whereas the dashed line is the asymptotic result. 

As seen above, the slope of the energy is the force required to deform the droplet. We therefore see that increasingly more force is required to compress the droplet up to the critical conformation, but that beyond the transition point the more we compress the droplet the less force is required to deform it. The non-monotonic force profile is  reminiscent of the non-monotonic pressure required to inflate a balloon \cite{muller02}. For  squeezed droplets, that result  means  that the droplet will buckle at the transition point if an increasing (or constant) load is applied. Unlike the classic Euler-buckling of beams \cite{landau86} which can support an increasing load after a buckling event, the buckling instability of a droplet is catastrophic, a so-called limit-point instability, as the droplet can no longer sustain the force at the point of instability and collapses.  In the context of superhydrophobic surfaces in which a droplet rests on a series of posts this would be what has been referred to as the  impalement transition between the Cassie and Wenzel states  \cite{quere05,bocquet11}. Precisely the same behavior will be obtained in three dimensions.

%%%%%%%%%
%%%%%%%%%
%%%%%%%%%

\subsubsection{Inclined surface}

How different is the transition to an asymmetric shape is the  upper surface is not flat? Since we do not want to embed a broken-symmetry in the top surface we insist that the top surface remains axisymmetric, the simplest example of which is a  cone of slope $\chi$ -- or a wedge in two dimensions. One finds that a droplet undergoes a similar symmetric to asymmetric bifurcation of possible solutions precisely at the moment when the droplet on the bottom surface (at $t=t_1$) is parallel to the upper surface, in other words, when the apparent contact angle with the bottom (pinned) surface reaches the value $\beta=\pi-\chi$. 
As we show below, this two-dimensional criterion is no longer indicative of the onset of the instability in three dimensions where the critical compression becomes a more complex function of the droplet volume and contact angle. The quantitative agreement between two and three dimensions is therefore restricted to the case where the top surface is flat.

%%%%%%%%%
%%%%%%%%%
%%%%%%%%%

\subsection{Three-dimensional analysis}
\subsubsection{Axisymmetric solution}
In three dimensions the problem becomes more complicated due to the introduction of an second radius of curvature. Unlike the two-dimensional case, asymmetric analytical extremum are unavailable. We follow here the approach of Russo and Steen  \cite{russo86} by first assuming that the droplet is axisymmetric, which yields analytical solutions. We then do  small asymmetric perturbations to the axisymmetric shapes to find the configuration where the axisymmetric solution is no longer energy minimizing. Finally in Section \ref{se} we use Surface Evolver\cite{brakke92} simulations to explore the mechanical properties of the asymmetric shapes.

The energy functional, Eq.~\eqref{energy}, may be written
\begin{align}
\Ef=\int_{S_{lv}}\gamma dA-\int_{S_{sl}}\gamma\cos\theta dA- \int_{V_l} p dV.
\end{align}
\begin{figure}
\includegraphics[scale=.85]{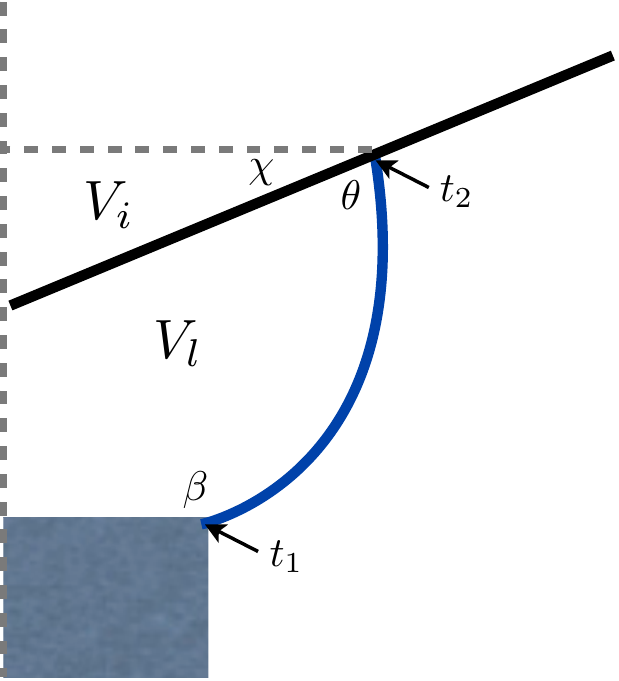}
\caption{Schematic representation of an axisymmetric pinned droplet compressed by a cone of slope $\chi$. The angle $\theta$ indicates the contact angle with the top surface ($t=t_2$) and $\beta$ the apparent contact angle with the bottom surface ($t=t_1$). Here $V_l$ is the volume of the liquid and $V_i$ is the volume of revolution inside the cone from its tip to $t_2$.}
\label{anglefig}
\end{figure}
The axisymmetric droplet is schematically depicted in Fig.~\ref{anglefig}. 
It is easier to integrate $V = V_l+V_i$ and so we have to subtract the volume $V_i$. A known function $g(r,z)=0$ defines the topology of the top surface. Assuming axisymmetry and parameterizing the shape  by $t$ we may recast the integral as
\begin{align}\label{e_3D}
\Ef=& \ \int_{t_1}^{t_2}\pi\left(2\gamma r\sqrt{\rd^2+\zd^2}-p r^2\zd \right) dt\nonumber\\
&-\gamma\cos\theta S_{sl}(r_2)+pV_i(r_2),
\end{align}
or simply $\Ef=\int_{t_1}^{t_2} F(r,\rd,\zd,t) dt+f(r_2)$, where again numbered subscripts indicate evaluation at $t_1$ or $t_2$ and the over-dot represents a partial derivative with respect to $t$. Note that since the topology of the upper surface is known then the only unknown for $S_{sl}$ and $V_i$ is  the location of the contact point, $r_2$. 

Extremizing Eq.~\eqref{e_3D} leads to the Young-Laplace equation
\begin{align}
\frac{d}{dt}\left[\frac{2\gamma r \zd}{\sqrt{\rd^2+\zd^2}}-p r^2\right]=0,
\label{laplace}
\end{align}
and the boundary condition
\begin{align}\label{bc}
\l.\l(\frac{\partial f}{\partial r_2}+\frac{\partial F}{\partial\rd}-\frac{\partial F}{\partial\zd}\frac{\partial g/\partial r}{\partial g/\partial z}\r)\r|_{t=t_2}=0.
\end{align}

Eq.~\eqref{laplace} was shown by Howe  \cite{howe87}, and restated in English by Gillette and Dyson  \cite{gillette71}, to have a solution
\begin{subeqnarray}\label{gensol}
r(t)&=\Lambda \sqrt{1-\sin^2\Omega \sin^2 t},\\
z(t)&=\Lambda[E_2(\Omega,t)+E_1(\Omega,t)\cos\Omega],
\end{subeqnarray}
where
\begin{align}
E_i=\int_{0}^{t}(1-\sin^2\Omega\sin t')^{-3/2+i}dt',
\end{align}
are elliptic integrals of the first and second kind. 
The solution described by Eq.~\eqref{gensol} has two parameters: the first one, $\Lambda$, with units of length, merely acts as a scaling factor; the second parameter, $\Omega$, which is dimensionless, modulates the shape, and may be interpreted through the mean curvature $H$ as follows
\begin{align}
\cos\Omega=\frac{1-\Lambda H}{\Lambda H}\cdot
\end{align}
From this point on we rescale all distances by $\Lambda$ (equivalently we set $\Lambda=1$). Note that using these parameters, the  pressure is then given by
\begin{align}
\frac{p}{\gamma}=\frac{2}{\Lambda(1+\cos\Omega)}\cdot
\end{align}
We note that the energy functional in Eq.~\eqref{e_3D} may be recast using a unit speed parameter $s$ (arclength of the generatrix) \cite{myshkis87} which then simplifies the formulae shown here; however, because the analytical axisymmetric solution is nonunit speed\cite{slobozhanin97,myshkis87} we leave the formulation general.

\subsubsection{Boundary conditions}
If the upper surface is a cone with slope $\chi$ as in Fig.~\ref{anglefig} then it is described by
\begin{align}
g(r,z)=r\tan\chi-(z-h)=0,
\end{align}
where $h$ is the smallest separation between the pinned surface and $g$. The flat wall is included here as the special case $\chi=0$. We now have
\begin{align}\label{above}
f(r_2)=-\gamma\cos\theta \frac{\pi r_2^2}{\cos\chi}+p\frac{\pi r_2^3}{3}\tan\chi.
\end{align}
Using Eq.~\eqref{above} into Eq.~\eqref{bc} we obtain the contact angle condition
\begin{align}\label{bccone}
\zd_2=\rd_2\tan(\theta+\chi).
\end{align}

By differentiating our solution, Eq.~\eqref{gensol}, and substituting into Eq.~\eqref{bccone} we can then obtain the location of the contact point as
\begin{align}
t_2=\frac{1}{2}\left(\theta +\chi -\arccos\left[-\cos(\theta +\chi)\cot\left(\frac{\Omega }{2}\right)^2\right]\right).
\end{align}
Note that if we were to select $\theta=\pi/2-\chi$ then we obtain $t_2=0$ as expected. Also since the droplet shape is symmetric about the point $t=0$ then the starting point (pinned contact point on the bottom surface) is given by the negative of the formula for $t_2$, and we have
\begin{align}
t_1=\frac{1}{2}\left(-\beta+\arccos\left[-\cos\beta\cot\left[\frac{\Omega }{2}\right]^2\right]\right),
\label{apparent}
\end{align}
where in Eq.~\eqref{apparent} $\beta$ is  the  apparent contact angle with the horizontal bottom surface. The point where $\beta=\pi$ is referred to as the Steiner limit\cite{russo86}. In the two-dimensional analysis, recall that we obtained that the droplet goes unstable when the apparent contact angle, $\beta$, is parallel to the upper surface and hence $t_1=t_1^c$ when $\beta=\pi-\chi$. We show below that this is true only when the wall is flat $\chi=0$, but does not remain valid when $\chi \neq 0$.

\subsubsection{Physical contact}
If the slope of the cone is nonzero, $\chi\ne0$, we run the risk of having physical contact between the upper surface and lower surface prior to the instability hence we have to limit the regime of $\Omega$ and $t_1$ (or $\beta$) to exclude this possibility. Since the parameter $\Lambda$ merely scales the shape it is inconsequential. Solving the equation
\begin{align}
z(t_2,\Omega)-z(t_1,\Omega)=r(t_2,\Omega)\tan\chi,
\end{align}
yields the limit of physically realizable solutions in the $\Omega$-$t_1$ plane.

\begin{figure*}
\includegraphics[scale=0.55]{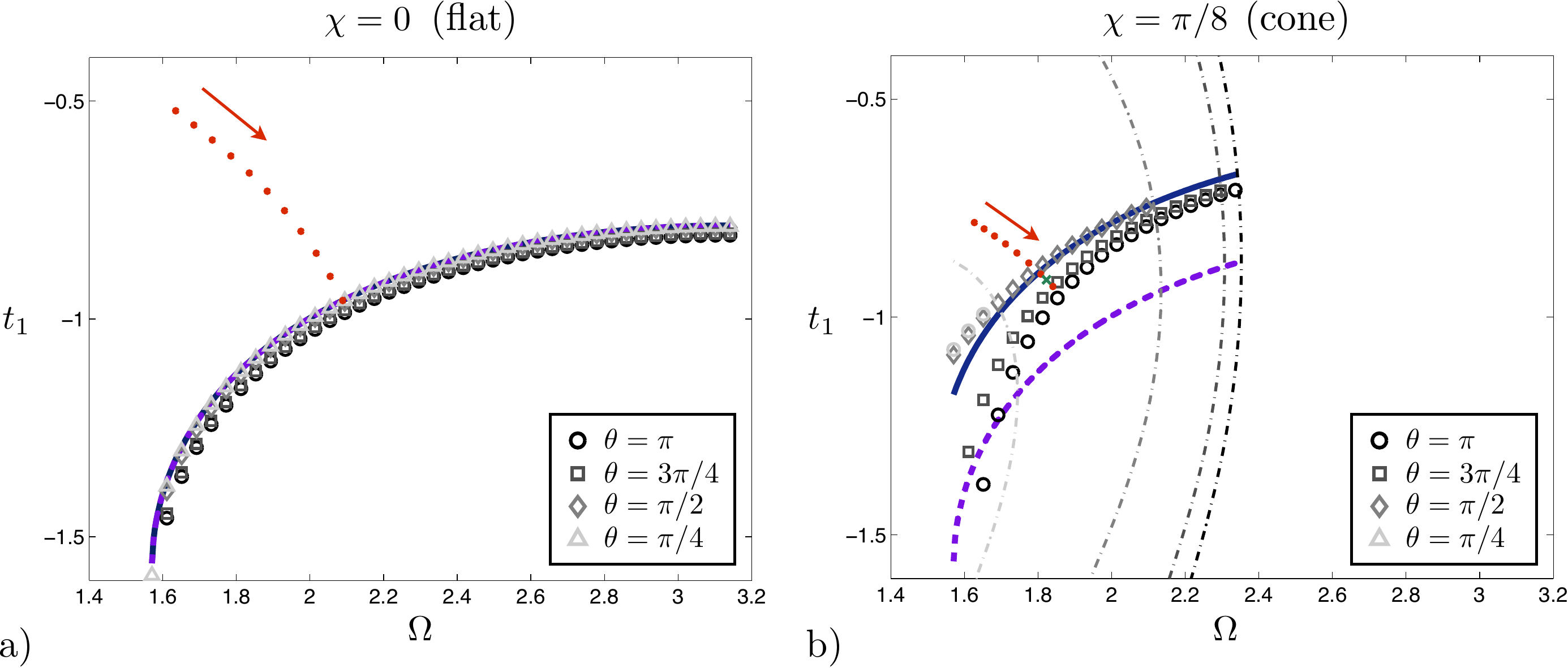}
\caption{The onset of the asymmetric instability with varying shape $\Omega$. The data points in this figure indicate the limit of stability found by solving $\Delta E(t_1)=0$ for a flat wall (case a, $\chi=0$) and a  conical top surface (case b,  $\chi=\pi/8$). 
The  contact angle $\theta$ on the top surface varies between $\pi/4$ and $\pi$ by $\pi/4$ increments.  The dashed line (purple online) indicates the point when the apparent contact angle $\beta=\pi$ (droplet tangent to the bottom surface) while the solid line (blue online) indicates when $\beta=\pi-\chi$ (bottom contact line parallel to the top surface); these two limits coincide in the flat wall case of (a). The dash-dot lines indicate the limit of physically realizable solutions when $\chi\ne0$. Our parameter $\alpha\in(-0.06,-0.29)$ is selected to give the earliest possible instability. We see when $\chi=0$ (flat wall) the limit of stability corresponds always to $\beta=\pi$, but when $\chi\ne0$ the data does not always lie on either the $\beta=\pi-\chi$ or $\beta=\pi$ curves. The dotted lines (red online) in each plot are examples of a paths of constant volume with arrows indicating the direction of decreasing gap separation; they are further illustrated  in Figs.~\ref{evolverdouble} and \ref{evolverdoublecone}.}
\label{separate}
\end{figure*}

\subsubsection{Perturbations}
In order to observe the instability of  axisymmetric shapes, following the method of Russo and Steen \cite{russo86}, we add to our axisymmetric solution (defined by $r(t)$ and $z(t)$) small, non-axisymmetric perturbations defined by the functions $F(t,\phi)$ and $G(t,\phi)$ which contain explicit dependence on the azimuthal angle $\phi$. The liquid-vapor interface is thus given by the following equations
\begin{align}
R(t,\phi)&=r(t)+F(t,\phi),\\
Z(t,\phi)&=z(t)+G(t,\phi).
\end{align}
The perturbations given by $F$ and $G$ must yield shapes which satisfy the boundary conditions and preserve volume, $\Delta V_l\equiv V_l(R,Z)-V_l(r,z)=0$, in order to be viable.

To discover the onset of the instability we solve for $t_1^c(\Omega)$ such that surface energy of the perturbed shape is equal to the axisymmetric shape $\Delta E=E(R,Z)-E(r,z)=0$. This will yield the point of the instability for any axisymmetric shape $\Omega$. The results of this perturbative analysis are given below while the mathematical details are given in Appendix \ref{AppA}. The perturbation analysis can only reveal the onset of the asymmetry; to probe the mechanical properties of the asymmetric droplets Surface Evolver simulations are used in Section \ref{se}. Alternatively, other schemes \cite{slobozhanin97,lowry00}, often developed for liquid bridges, might be adapted for use with the contact angle boundary condition.

%%%%

\subsubsection{Results}

Our theoretical results are illustrated in  Fig.~\ref{separate}. We plot the dependance of the critical point at which  the axisymmetric state becomes unstable to non-axisymmetric perturbations, $t_1^c(\Omega)$, with the shape parameter, $\Omega$,  for contact angles  on the upper surface ranging from $\pi/4$ to $\pi$. The case of a flat wall is shown in Fig.~\ref{separate}a while the conical surface with $\chi=\pi/8$ is displayed in Fig.~\ref{separate}b. In both figures, the dotted lines (red online) are examples of a paths of constant volume with arrows indicating the direction of decreasing gap separation.

Plotted in Fig.~\ref{separate} as a dashed line (purple online) is the Steiner limit which corresponds to an apparent contact angle on  the (bottom) pinned surface of $\beta=\pi$, while the solid line (blue online) indicates where the apparent contact angle at the bottom surface is parallel to the top surface, i.e. $\beta=\pi-\chi$. In the case of a flat wall, $\chi=0$, the two lines are coincident. For this geometry we see that for any contact angle on the upper surface the point of instability occurs on or immediately after the Steiner limit. Since we use only a truncated and restricted series to represent the asymmetric perturbations (see Appendix \ref{AppA}), this results represents an upper bound.
Using a symmetrization argument, Gillete and Dyson showed that the droplets in a liquid bridge\cite{russo86,meseguer95,lowry95,slobozhanin97,lowry00}, which can be represented by single valued functions $r(z)$ (in other words droplets prior to the Steiner limit), are stable to axisymmetric perturbations\cite{gillette72}. There is nothing in their argument that prohibits its application when we permit $r_2$ to be variable so long as $z_2$ is fixed; however, such a map does not preserve contact angle. To apply in our case when $\chi=0$ we must adjust the contact angle, but it can easily be shown that one can always smoothly change the contact angle with a vanishing change in the surface area and volume. As lower and upper bound coincide, we get the final result that, in the flat wall case, the limit of stability is when the droplet is parallel to the pinned surface at the point of contact (apparent contact angle of $\pi$), similarly to the stability threshold for liquid bridges and sessile droplets \cite{michael81,lowry00}.

The case where the top surface is conical ($\chi=\pi/8$, Fig.~\ref{separate}b) shows a different stability behavior.  Because the upper surface may now contact the lower surface before the droplet goes unstable, a portion of parameter space, depending on the contact angle, is excluded; this is indicated by the dash-dotted lines. Our stability calculations show, in this case, that  the droplet does not always go unstable when its surface is parallel to the upper surface (blue solid line) nor when it is tangent to the lower surface (purple dashed line). The two-dimensional prediction cannot therefore be extended to three dimensions. 

Unlike, for the flat surface, the symmetrization argument clearly does not hold when $\chi\ne0$ and hence our prediction here for the limit of stability can only be regarded as an upper bound. We will show in the next section that this prediction can perform well nevertheless.

%%%%%%%%%%%%%%%%%%%%%%%%%%%
%%%%%%%%%%%%%%%%%%%%%%%%%%%
\section{Surface Evolver computations}\label{se}

In order to corroborate our asymptotic predictions we appeal to numerical simulations using  Surface Evolver (SE) \cite{brakke92}. Nagy  and Neitzel also used the same program to confirm the existence of their observed instability in an idealized setting \cite{nagy09}. In Surface Evolver the droplet shape is discretized into triangular facets whose positions are defined by a position vector $\bX$ hence the energy is $E(\bX)$ \cite{brakkeweb}. At a minimum of energy we have $\bnabla E=0$ and the Hessian matrix, $\bH=\partial^2E/\partial X_i \partial X_j$,  must be positive definite. We run SE simulations of both the flat and conical geometries and see that, as the separation $h$ approaches a critical separation $h^c$, the smallest eigenvalue of $\bH$ approaches zero. If the separation is reduced below this critical separation then the symmetric form yields negative eigenvalues and the true minimum is then given by an asymmetric shape.

\begin{figure}
\includegraphics[scale=0.55]{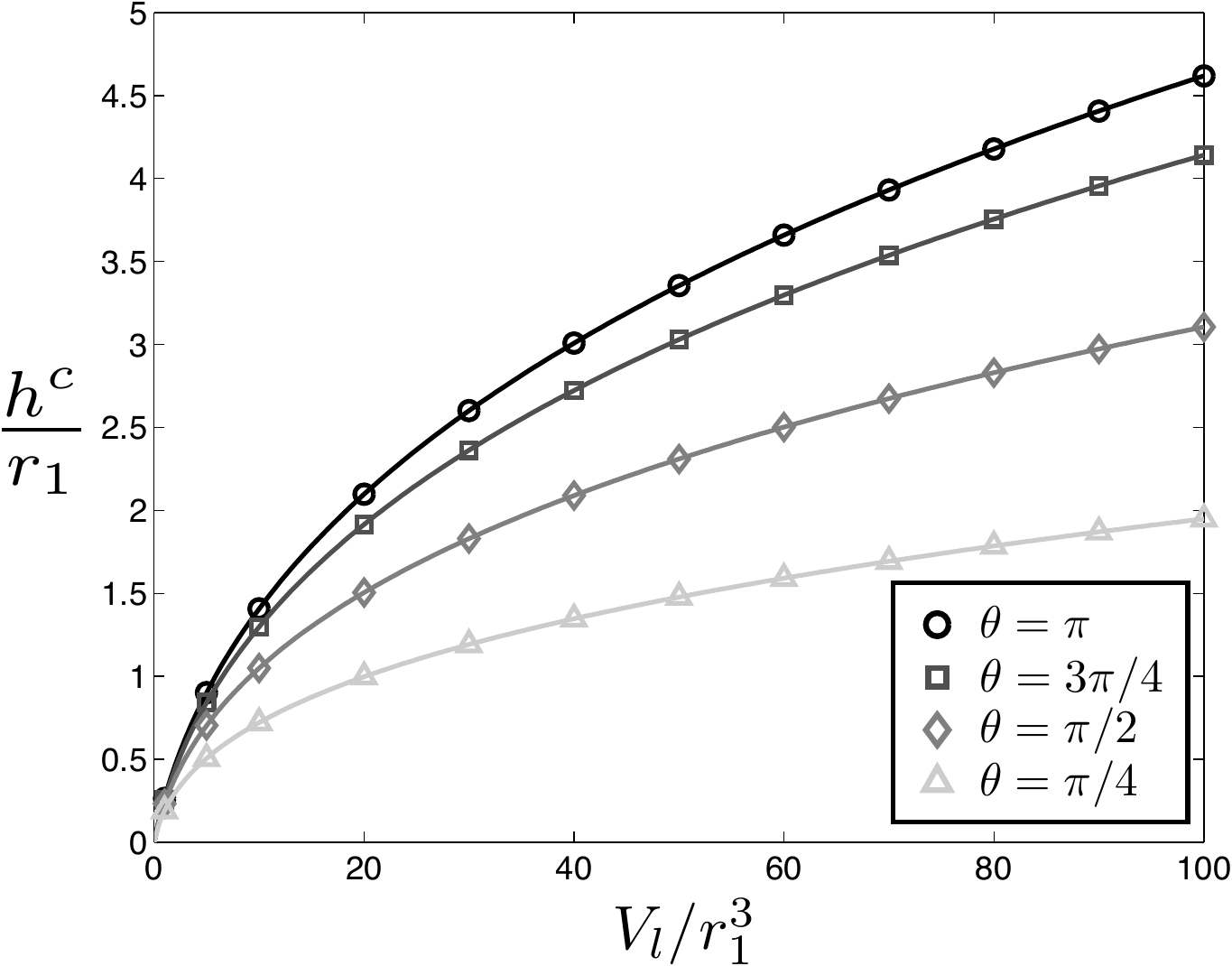}
\caption{Critical separation $h^c$ vs liquid volume $V_l$ non-dimensionalized by the pinned radius $r_1$ for $\theta=\{\pi/4,\pi/2,3\pi/4,\pi\}$ and $\chi=0$.}
\label{stability}
\end{figure}

Appealing to dimensional analysis tells us that we can write the critical separation in the form $h^c/r_1=\Phi(V_l/r_1^3,\theta,\chi)$ \cite{russo86,barenblatt96}. Given that we have an analytical equation for the the limit of stability when the compressing top wall is flat ($\chi=0$) in Eq.~\eqref{apparent}, we can plot this function easily for various constant $\theta$, with results shown in solid lines in Fig.~\ref{stability}. We then run Surface Evolver, the symbols in Fig \ref{stability} indicate simulations performed where $h=h^c$ was stable but $h=0.99h^c$ unstable. This then demonstrates that the instability occurs precisely as predicted theoretically in the previous section. We note that, as predicted by the two-dimensional example, the instability is the least physically apparent for large volumes and high contact angles, $\theta\rightarrow \pi$. In Surface Evolver this manifests itself by a need to use very high resolution to properly capture the onset of the stability in these physical regimes. 

\begin{figure*}
\includegraphics[scale=0.42]{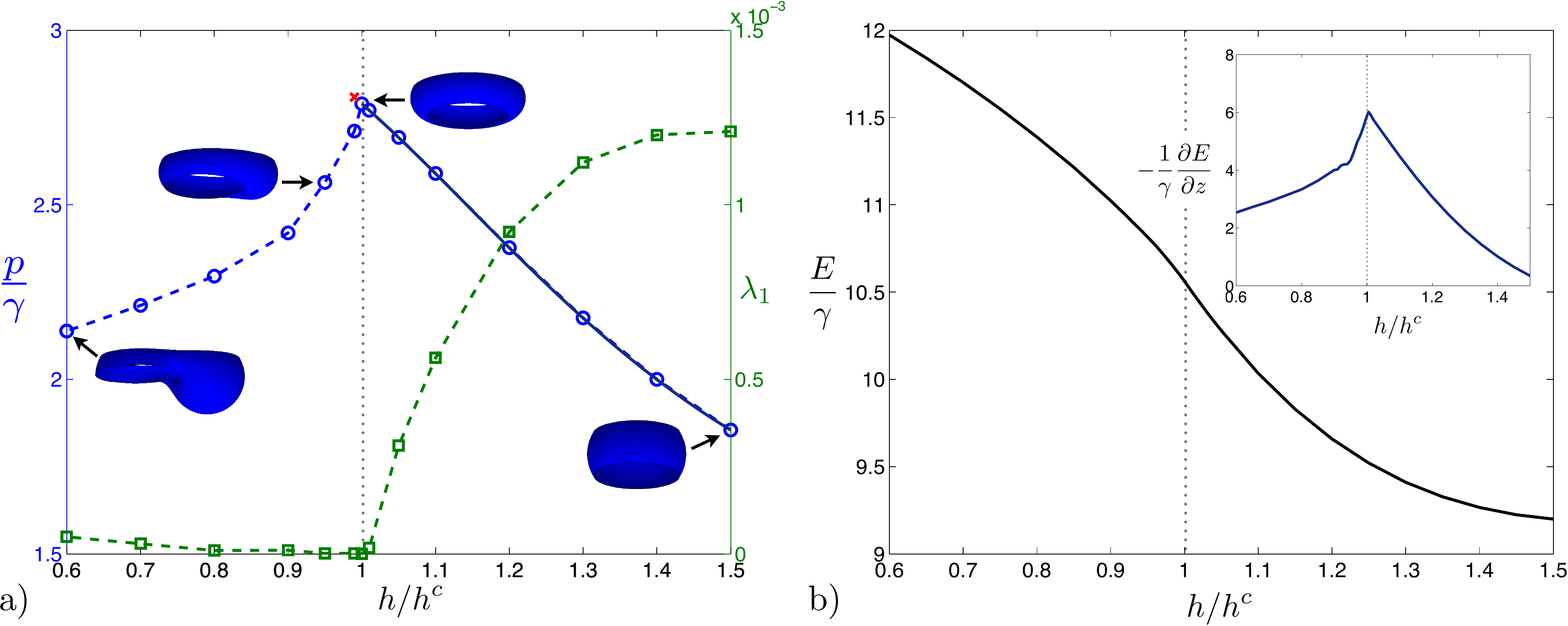}
\caption{Surface evolver simulations of the onset of the instability for a flat wall, $\chi=0$ with $V=4$ and $r_1=1$. (a): Pressure, $p$ (axis on the left, circles joined by a dashed line is numerical data and the solid line is analytical), and smallest eigenvalue of the Hessian matrix, $\lambda_1$ (squares with axis on right), versus separation $h$. We see that the smallest eigenvalue goes to zero, indicating a loss of stability, precisely when we predict $h=h^c$. The pressure reaches a maximum at the onset of the shape instability. The cross (red online) indicates the pressure past $h^c$ is higher if the conformation were axisymmetric but this point is not physically realized. 
We have also inset graphical representations of the droplet for $h/h^c=0.6,0.95,1,$ and $1.5$. (b): Surface energy, $E$, varies monotonically with the gap height. The slope of the energy (inset) indicates that the force required to deform the droplet peaks at $h=h^c$ past which the droplet becomes asymmetric and the amount of force required to deform the droplet decreases with increasing displacement. A loaded droplet would thus buckle at the transition to asymmetry.}
\label{evolverdouble}
\end{figure*}

As a more concrete example we define a specific physical system and illustrate its transition to asymmetry. We choose as an example a droplet of volume $V=4$ pinned on a base of radius $r_1=1$ (arbitrary units) compressed by a flat upper surface of contact angle $\theta=120^\circ$. Solving for $V(\Lambda,\Omega,t_1^c(\Omega))=4$ and $r(\Lambda,\Omega,t_1^c(\Omega))=1$ we determine theoretically that $\Lambda\approx 1.42$ and $\Omega\approx 2.089$ at a predicted critical separation $h^c\approx0.6923$. The constant volume path of this system, from $h=1.5h^c$ to $h^c$, was illustrated in parameter space in Fig.~\ref{separate} by dots (red online), the arrow indicating the direction of increased compression of the droplet.

In Fig.~\ref{evolverdouble} we show the results of Surface Evolver simulations of this system with the separation, $h$, ranging $1.5h^c$ to $0.6h^c$. First and foremost, we see  that the instability occurs precisely at the analytically predicted $h=h^c$. The pressure in the droplet increases until the critical point where the smallest eigenvalue in the system vanishes, indicating a loss of stability. Beyond the critical point, symmetric shapes have negative eigenvalues indicating an unstable saddle. The stable shapes beyond the critical point are asymmetric and have progressively lower pressure. In Fig.~\ref{evolverdouble}b we display  the variation of the  surface energy of the droplet, $E$, with the separation distance between the surface; in inset we plot the slope of the energy (obtained by numerical differentiation of $E$), i.e.~the force acting on both surfaces resisting compression. The surface energy progressively increases as we deform the droplet from $h=1.5h^c$ to $h=0.6h^c$.  The slope of the energy (shown in the inset) indicates that the force required in order to further deform the droplet peaks at the critical separation and then monotonically decreases with increasing deformation. 

Experimentally, if we were to progressively load a droplet with increasing force then at the point when the droplet is tangent to the pinned area it would become mechanically unstable, regardless of the contact angle of the deforming surface (and true for both two- and three-dimensional systems). The system would undergo a dynamic collapse past this point. These mechanical considerations are important for the design of systems such as the one described by Neitzel et al.~for load support in a non-wetting scenario  \cite{neitzel02, aversana04, nagy09}. The peak load occurs at the critical shape, beyond which the droplet undergoes a limit-point buckling instability, and collapses.

\begin{figure}[t!]
\includegraphics[scale=0.42]{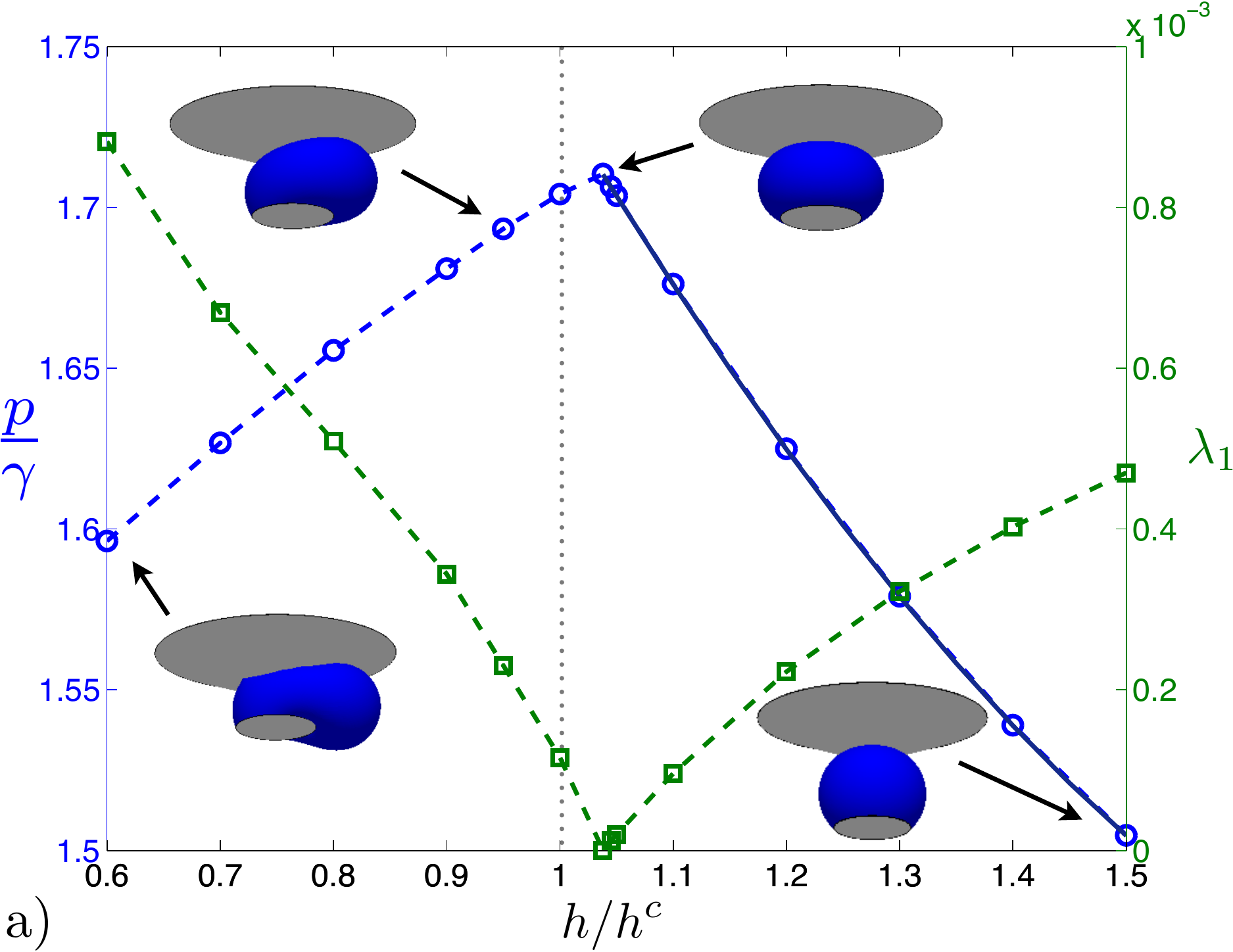}
\caption{Surface Evolver simulations of the onset of the instability in the case where the top surface is conical, $\chi=\pi/8$. Same notation and parameters  as in Fig.~\ref{evolverdouble}.
The smallest eigenvalue goes to zero, and the  pressure reaches a maximum, just past our calculated stability limit, at about $h\approx 1.04h^c$. Insets show graphical representations of the droplet for $h/h^c=0.6,0.95,1$ and $1.5$.}
\label{evolverdoublecone}
\end{figure}

With $\chi\ne0$ we are not assured that an axisymmetric instability can occur prior to contact between the upper and lower surface.  Since we do not have a analytical expression for the unstable point, when we must solve $\Delta E(\Omega,t_1^c)=0$ given by Eq.~\eqref{dF}, together with $V(\Lambda,\Omega,t_1^c)=10$ and $r(\Lambda,\Omega,t_1^c)=1$, with $\chi=\pi/8$ and $\theta=3\pi/4$, to obtain our predictions $\Lambda\approx 1.575$, $\Omega\approx 1.84$ and a contact point $t_1^c\approx -0.93$. We then compute our predicted critical separation $h^c=z_2-z_1^c-r_2\tan\chi\approx 1.18$. As discussed the analytical prediction serves as an upper bound on the energy at the critical point, and we see when compared to the SE simulations, shown in  Fig.~\ref{evolverdoublecone}, that the analysis slightly under-predicts the critical separation, and we obtain numerically that the droplet becomes asymmetric when $h\approx 1.038h^c$. 

The path from $h=1.5h^c$ to $h=0.6h^c$ was shown earlier in Fig.~\ref{separate}b) by dotted lines (red online), the cross (green online) along this path indicates where the instability occurs in Surface Evolver. Similarly to the flat wall case, we find the droplet displays a drop in pressure after losing axisymmetry indicating the presence of a limit-point buckling instability.

%%%%%%%%%%%%%%%%%%%%%%%%%%%
%%%%%%%%%%%%%%%%%%%%%%%%%%%
\section{Conclusion}
 
In this paper, motivated by recent experiments by Nagy and Neitzel \cite{nagy09}, we have used theory and computations to show that if a  droplet, pinned at the bottom by a surface of finite area,  if sufficiently deformed by a  surface at the top, will always develop a shape instability at a critical compression - a result true for all values of the contact angle between the droplet and the top surface. After the critical compression, the droplet will then transition from a symmetric shape to an asymmetric shape. The force required to deform the droplet peaks at the critical point then progressively decreases indicative of a buckling instability. If the deforming surface is flat then we predict the instability to occur when the apparent contact angle of the droplet at the pinned surface is $\pi$, regardless of the contact angle of the upper surface, similarly to past work on liquid bridges and sessile droplets. However, when the upper surface has non-trivial topology this criterion no longer holds, and a detailed stability analysis is carried out to predict the critical compression. An interesting question for future work would be to explore the effects of  surface curvature on the shape instability.

\begin{acknowledgments}
Funding by the NSF (CBET-0746285) and NSERC (PGS D3-374202) is gratefully acknowledged.
\end{acknowledgments}

\appendix

\section{Perturbations}\label{AppA}

To satisfy the requirements on the functions $F$ and $G$ we expand them in a quarter range basis as
\begin{align}
F(t,\phi)&=\sum_{m,n}A_{m,n}\cos(m\phi)\sin\l(\frac{n\pi}{2}T\r),\\
G(t,\phi)&=\sum_{m,n}B_{m,n}\cos(m\phi)\sin\l(\frac{n\pi}{2}T\r),
\end{align}
where $n$ is odd and $T=(t-t_1)/\Delta t$.

Using the above definitions of $F$ and $G$ in Eq.~\eqref{G2F2} we obtain
\begin{align}\label{bcpert2}
\sum_n(-1)^{\frac{n-1}{2}}B_{m,n}&=\sum_n (-1)^{\frac{n-1}{2}} A_{m,n}\tan\chi.
\end{align}
Unlike the analysis in Russo and Steen  \cite{russo86} we cannot use normal perturbations with zero magnitude at the endpoints as we wish for the droplet to be able to have a variable contact point on the upper surface (fixed contact angle condition). However, as shown in Ref.~\cite{russo86} the lowest modes are the ones with the largest increase in surface energy, therefore in order to establish an upper bound on the energy at the critical point we restrict our analysis to  the lowest two modes in $t$ and $\phi$.

We first consider the case $\chi=0$, then generalize. With Eq.~\eqref{bcpert2} we have $B_{m,1}=B_{m,3}$. The boundary condition places no restriction on $g$ and hence we let $A_{m,3}=0$. As the surface energy is nonlinear in the shape perturbations,  different modes couple and  we must be careful in our selection of allowed perturbation shape. While we cannot make the perturbations normal to the shape everywhere, we impose a related constraint by letting $B_{m,1}=\alpha A_{m,1}$, where $\alpha$ is a free parameter. This  restriction on the shapes reduces the unknown coefficients by one to facilitate the analytical calculation, while the free parameter allows  some flexibility on the shape of the perturbations, and we tune its value so as to give the earliest possible instability. Because our choice for the space of allowed perturbations may not optimally minimize the surface energy, our analysis is thus only able to derive an upper bound for the stability limit. However, as we will show, this upper bound  will coincide with the lower bound, when $\chi=0$.
With these assumptions,  our perturbations now take the form
\begin{align}
F(t,\phi)&=(A_{01}+A_{11}\cos\phi)f(t),\\
G(t,\phi)&=(A_{01}+A_{11}\cos\phi)g(t),
\end{align}
where
\begin{align}
f(t)&=\sin\l(\frac{\pi}{2}T\r),\\
g(t)&=\alpha\l[\sin\l(\frac{\pi}{2}T\r)+\sin\l(\frac{3\pi}{2}T\r)\r].
\end{align}

For a general  value of $\chi\neq 0$, we can rotate our perturbations to the $\chi$ plane $[f',g']^T= \bR_\chi[f,g]^T$ where $\bR_\chi$ is a two-dimensional rotation operator of angle $\chi$. Hence, for all $\chi$, we may write
\begin{align}
f(t)&=\l[(\cos\chi-\alpha\sin\chi)\sin\l(\frac{\pi}{2}T\r)-\alpha\sin\chi\sin\l(\frac{3\pi}{2}T\r)\r],\\
g(t)&=\l[(\alpha\cos\chi+\sin\chi)\sin\l(\frac{\pi}{2}T\r)+\alpha\cos\chi\sin\l(\frac{3\pi}{2}T\r)\r].
\end{align}
We see that $G_2=F_2\tan\chi$ for all $\alpha$.

We require the asymmetric shapes to satisfy the contact angle condition at $t_2$ and hence $\Zd_2=\Rd_2\tan(\theta+\chi)$ which leads directly to
\begin{align}\label{bcpert}
\Gd_2&=\Fd_2\tan(\theta+\chi).
\end{align}
Eq.~\eqref{bcpert} merely states that our asymmetric perturbations must preserve the slope imposed by the contact angle at $t_2$. To satisfy this for all $\theta$ and $\chi$ we set $\Fd_2=\Gd_2=0$. Furthermore we require $F_1=G_1=0$ so that the droplet remains pinned at $t_1$. Additionally, our perturbations must be directed along the surface at the upper bound and hence we need 
\begin{align}\label{G2F2}
G_2=F_2\tan\chi.
\end{align}

We now expand $F=\sum\epsilon^j F^{(j)}$ and $G=\sum\epsilon^j G^{(j)}$ where $\epsilon$ is a small dimensionless parameter. We must ensure our perturbations, order by order, conserve volume
\begin{align}\label{volvol}
V_l=\frac{1}{2}\int_{t_1}^{t_2}\int_{0}^{2\pi}R^2\Zd dt d\phi-\frac{1}{6}\tan\chi\int_{0}^{2\pi}R_2^3 d\phi,
\end{align}
where the second term in Eq.~\eqref{volvol} takes into account the inner cone.  Expanding and subtracting off the unperturbed case we get to leading order
\begin{align}
\Delta V_l=\epsilon\pi  A_{01}^{(1)}  \l[\int_{t_1}^{t_2} r \left(r\gd+2 f \zd\right)dt-f_2r_2^2\tan\chi\r]=0,
\end{align}
hence $A_{01}^{(1)}=0$. The $\Of(\epsilon^2)$ term yields the following relationship
\begin{align}
A_{01}^{(2)}=-\l(A_{11}^{(1)}\r)^2\frac{ \int_{t_1}^{t_2}f \left(2r\gd+f \zd\right)dt-\frac{1}{2}f_2^2r_2\tan\chi}{2\int_{t_1}^{t_2}r\left(r\gd+2 f \zd\right)dt-f_2r_2^2\tan\chi},
\label{quo}
\end{align}
hence with $A_{01}^{(2)}=-\l(A_{11}^{(1)}\r)^2 I$, where $I$ is the quotient in Eq.~\eqref{quo}, volume is conserved to order $\Of(\epsilon^2)$.

The perturbed surface energy is given by
\begin{align}
E=&\gamma \int_{t_1}^{t_2}\int_{0}^{2\pi}\l[ R^2(\Rd^2+\Zd^2)+(R_\phi \Zd - \Rd Z_\phi)^2 \r]^{1/2} dtd\phi\nonumber\\
 &-\frac{\gamma\cos\theta}{2\cos\chi}\int_{0}^{2\pi}R_2^2d\phi.
\end{align}
Expanding the integral and subtracting off the axisymmetric contribution, the leading order term of the difference is given by
\begin{align}\label{dF}
\frac{\Delta E}{\epsilon^2\pi\gamma\l(A_{11}^{(1)}\r)^2}=&\int_{t_1}^{t_2}\bigg\{-2If v-\frac{rw^2}{2 v^3}+\frac{1}{2 rv}\Big(2rw(f-2Ir)\nonumber\\
&+r^2 \left(\fd^2+\gd^2\right)+\left(g \rd-f \zd\right)^2\Big)\bigg\}\nonumber\\
&-\frac{1}{2}\frac{\cos\theta}{\cos\chi}f_2(f_2-4Ir_2),
\end{align}
where
\begin{align}
v(t)&=\sqrt{\rd^2+\zd^2},\\
w(t)&=\fd \rd+\gd\zd.
\end{align}
Note from Eq.~\eqref{dF} that our marginal stability curve will be independent of $\gamma$ and $A_{11}^2$. We are interested in all outward bulging shapes, nodoids, and hence we check all $\Omega\in[\pi/2,\pi)$.

\bibliography{droplet}

\end{document}